\documentclass[journal,onecolumn,12pt]{IEEEtran}

\usepackage{amsmath}
\usepackage[dvipsnames]{xcolor}

\usepackage{txfonts}
\usepackage{tikz}
\usetikzlibrary{arrows}

\usepackage{multirow}

\newtheorem{theorem}{Theorem}
\newtheorem{corollary}{Corollary}
\newtheorem{definition}{Definition}
\newtheorem{example}{Example}

\usepackage{color}

\begin{document}

\title{Provision of Public Goods on Networks: On Existence, Uniqueness, and Centralities}

\author{Parinaz~Naghizadeh and 
        Mingyan~Liu%
\thanks{P. Naghizadeh and M. Liu are with the Department
of Electrical Engineering and Computer Science, University of Michigan, Ann Arbor, MI. e-mail: \{naghizad, mingyan\}@umich.edu}%
\thanks{A preliminary version of this work appeared in Allerton 2015 \cite{naghizadeh15}.}%
}

\markboth{Provision of Public Goods on Networks}%
{Naghizadeh and Liu}

\maketitle

\begin{abstract}
We consider the provision of public goods on networks of strategic agents. We study different effort outcomes of these network games, namely, the Nash equilibria, Pareto efficient effort profiles, and semi-cooperative equilibria (effort profiles resulting from interactions among coalitions of agents). We identify necessary and sufficient conditions on the structure of the network for the uniqueness of the Nash equilibrium. We show that our finding unifies (and strengthens) existing results in the literature. 
We also identify conditions for the existence of Nash equilibria for the subclasses of games at the two extremes of our model, namely games of strategic complements and games of strategic substitutes. 
We provide a graph-theoretical interpretation of agents' efforts at the Nash equilibrium, as well as the Pareto efficient outcomes and semi-cooperative equilibria, by linking an agent's decision to her centrality in the interaction network. Using this connection, we separate the effects of incoming and outgoing edges on agents' efforts and uncover an alternating effect over walks of different length in the network. 
\end{abstract}


\section{Introduction} \label{sec:intro}
\IEEEPARstart{W}e study strategic interactions in a network of agents who exert effort towards the provision of a public good. In these settings, the effort exerted by an agent affects not only herself, but also other agents interacting with her. This problem appears in  many social and economic applications. We present some applications. 

First, consider the \emph{spread of information and innovation} in networks.  New technologies developed by one entity/agent in the network may later be adopted by other agents in the network. The  interactions determining these innovation spillovers can in general depend on factors such as geographic location \cite{audretsch04} and the interacting agents' access to resources \cite{foster10}. Given this network, the possibility of spillovers can affect the decision of agents for investing in innovation or experimenting with new methods, leading to possible free-riding behavior. Specifically, a neighbor's effort can be either a \emph{substitute} or a \emph{complement} to an agent's own effort. Strategic substitutes (complements) are defined by the property that an increase of effort by an agent decreases (increases) her neighbors' marginal utilities, leading them to decrease (increase) their effort levels in response. 
For instance, if farmers in a village have the option of experimenting with a new variety of seeds, then those whose neighbors are experimenting are less likely to do so themselves \cite{foster10}. In this example, neighbors' efforts are a {substitute} to an agent's own effort. It may also be the case that an agent needs to increase her levels of experimentation in response to that of her neighbors, in order to remain competitive in her industry. In that case, the neighbors' efforts are a {complement} to the agent's own effort. 

Another setting of interest is \emph{investments in security} by interdependent entities. Security has been commonly viewed as a public good; examples include the model of airline security in \cite{kun03}, as well as the study of cyber-security in \cite{varian04,walrand11,miura08,grossklags08,naghizadeh16}. Investment in security by a neighbor can act as either a {substitute} or a {complement} to an agent's own effort. For example, in weakest target games \cite{grossklags08}, neighbors' efforts are complementary since the agent with the lowest security will be selected as the target by an attacker.  For total effort games \cite{varian04,grossklags08}, on the other hand, neighbors' efforts act as substitutes, as an agent's overall security is assumed to be determined by the sum of her own investment and her neighbors' efforts.  

In addition to the above applications, creation of new community parks or libraries in cities \cite{ballester10}, investment in pollution reduction measures by neighboring towns \cite{elliott15}, and even the states of happiness of individuals on a social network \cite{fowler08}, can be studied using this framework. 

The public good provision game studied in this paper belongs to the growing literature on  games on networks; see \cite{jackson14,bramoulle15} for recent surveys.  Specifically, we consider games in which, given the network structure, an agent's payoff depends on her own effort, as well as a \emph{weighted sum} of her neighbor's efforts. Our model allows for both complements and substitutes, different strengths of interactions (weighted graphs), and unidirectional interactions (directed graphs). We are interested in the study of Nash equilibria, Pareto efficient effort profiles, and semi-cooperative equilibria (we define these as the effort profiles emerging when coalitions of agents interact with one another). Our results provide an understanding of how the aforementioned outcomes (i.e., the results of agents' strategic interactions) are affected by the properties of the network. 

Our first result identifies necessary and sufficient conditions on the structure of the network (in terms of the dependence matrix) that guarantee that a Nash equilibrium exists and is \emph{unique}. We will show that previous results on the uniqueness of the Nash equilibrium \cite{naghizadeh15, miura08, ballester10, bramoulle14} can be recovered as corollaries of our first theorem. 

In addition to studying uniqueness, we identify (weaker) necessary and sufficient conditions for the \emph{existence} of Nash equilibria in two classes of games at the extremes of our model,  namely games with strategic complements and games with strategic substitutes. The identified conditions (for both existence and uniqueness) are solely based on the structure of the network. 

We then establish a connection between the agents' centrality in their dependence network, and the effort they exert at different interior Nash equilibria, Pareto efficient outcomes, and semi-cooperative equilibria. We separate the effects of dependencies (outgoing edges of the interaction network) and influences (incoming edges of the interaction network) on agents' effort decisions. We further discuss how the formation of coalitions is reflected in the centrality-effort characterization. We then uncover an \emph{alternating effect} along walks of different length in the network. We show that in a network with strategic substitutes, this alternating effect implies that changes along each walk of odd (even) length will negatively (positively) affect the agent's final decision. We provide additional intuition and examples for general networks in Section \ref{sec:cent-eff}.

\subsection*{Related literature}\label{sec:related}
Public good provision games, and network games in general, have recently received increasing attention. We refer the interested reader to \cite{jackson14,bramoulle15} for surveys on this general area. Here, we present the work most related to the current paper. 

Most of the existing work has studied the Nash equilibrium of network games. Previous work on identifying conditions for existence and uniqueness of Nash equilibria in public good provision games include \cite{naghizadeh15, miura08, ballester10, bramoulle14}. Both \cite{miura08,naghizadeh15} identify a similar sufficient condition for the existence of the Nash equilibria for public good provision games. The authors of \cite{bramoulle14} present a different sufficient condition for the uniqueness of the Nash equilibrium. Their result illustrates the role of the \emph{lowest} eigenvalue of the network in determining the outcome of strategic interactions. Finally, \cite{ballester10} provides necessary and sufficient conditions for the uniqueness of the Nash equilibrium in a class of games with hidden complementarities. In addition to identifying the necessary condition for uniqueness of Nash equilibria in general networks, we show that the sufficiency results of  \cite{naghizadeh15, miura08, ballester10, bramoulle14} can be recovered as corollaries of our main theorem. This comparison will further illustrate the key role of the lowest eigenvalue in (asymmetric) games with complementarities (in addition to the symmetric networks and particular classes of asymmetric networks studied in \cite{bramoulle14}). 

Our work is also closely related to \cite{elliott15,ballester06}, which provide graph-theoretical interpretations of agents' efforts in terms of their centralities in a suitably defined network. {The work of Elliott and Golub in \cite{elliott15} focuses mainly on the implementation of Pareto efficient outcomes. The current work and \cite{elliott15} differ in the network used as the basis of analysis: rather than working directly on the dependence matrix, \cite{elliott15} focuses on a \emph{benefits matrix} that is derived from the network graph; an entry $B_{ij}$ of the matrix is the marginal rate at which $i$'s effort can be substituted by the externality of $j$'s action. The authors show that Lindahl outcomes can be interpreted as node centralities in this benefits matrix.} 
Ballester et al. \cite{ballester06}, on the other hand, study the Nash equilibrium of a linear quadratic interdependence model, and relate the equilibrium effort levels to the nodes' Bonacich centralities in a suitably defined matrix of local complementarities. Despite the difference in the base models, both games have the same linear best-reply functions. As a result, the characterization of Nash equilibria based on Bonacich centralities (used in \cite{ballester06}) and alpha-centralities (used in this paper) are equivalent (see footnote \ref{f:bon-alpha}). We will see that using (the more general measure of) alpha-centrality allows us to provide graph-theoretical interpretations of Pareto efficient efforts and semi-cooperative equilibria as well. 

\subsection*{Main contributions}
The main contributions of this work are summarized as follows: 

$\bullet$ We identify the necessary and sufficient condition for \emph{uniqueness} of Nash equilibria in public good provision games. We show that our result unifies (and strengthens) previous results in the literature.  

$\bullet$ We identify the necessary and sufficient condition for the \emph{existence} of Nash equilibria in two subclasses of our model, namely games with strategic substitutes and games with strategic complements. 

$\bullet$ We present a \emph{graph theoretical characterization} of agents' actions at different effort outcomes, namely the Nash equilibria, Pareto efficient outcomes, and semi-cooperative equilibria (in terms of node centralities). Our characterization separates the effects of agents' dependencies and influences. It also uncovers an interesting alternating effect over walks of different length. 

\vspace{0.12in}

The remainder of the paper is organized as follows. We present the model for public good provision games in Section \ref{sec:model}, followed by conditions for the existence and uniqueness of Nash equilibria in Section \ref{sec:e&u}. Section \ref{sec:cent} discusses the graph theoretical characterization of different effort outcomes. In Section \ref{sec:coalitions}, we generalize the graph-theoretical characterization to games in which agents belong to different coalitions. Section \ref{sec:conclusion} concludes the paper.

\section{Model and Preliminaries} \label{sec:model}

\subsection{Public good provision games}

We study the strategic interactions of $N$ agents constituting the vertices of a directed network $\mathcal{G}=(\mathcal{N}, \mathcal{E})$; where $\mathcal{N}$ and $\mathcal{E}$ denote the set of agents and links, respectively. Each agent $i\in\mathcal{N}$ chooses to exert \emph{effort} $x_i\in \mathbb{R}_{\geq 0}$ towards the provision of a public good.\footnote{We follow Mas-Collel, Whinston, and Green \cite{mas-micro}, and define public goods as those that  are non-rivalrous; i.e., goods for which consumption by an agent does not reduce its availability to others. We therefore allow for both complements and substitutes, as well as both excludable and non-excludable public goods. We only explicitly make the distinction based on excludability in Section \ref{sec:coalitions}, when studying effort profiles that emerge under coalitions.} Agent $i$'s payoff depends on her own effort, as well as the effort exerted by other agents in her local neighborhood $\mathcal{N}_i:=\{j | \{i\rightarrow j\} \in \mathcal{E}\}$. An edge $\{i\rightarrow j\}$ indicates that agent $i$ \emph{depends} on agent $j$. The strength and type of this dependence are determined by the weight $g_{ij}\in \mathbb{R}$ of the edge $\{i\rightarrow j\}$. In particular, $g_{ij}>0$ ($<0$) indicates that $j$'s effort is a substitute (complement) to $i$'s effort. Let $\mathbf{G}=(g_{ij})$ denote the \emph{dependence} matrix of the graph. 

Let $\mathbf{x}=\{x_1, x_2, \ldots, x_N\}$ denote the profile of efforts exerted by all agents. 
The utility of agent $i$ at this effort profile is given by: 
\begin{align}
u_i(\mathbf{x}; \mathbf{G}) = b_i(x_i + \sum_{j\in \mathcal{N}_i} g_{ij}x_j) - c_ix_i~.
\label{eq:weff}
\end{align}
Here, $c_i>0$ is the marginal cost of effort for agent $i$, and $b_i(\cdot)$ is a twice-differentiable, strictly increasing, and strictly concave function, determining the benefit to agent $i$ from the aggregate effort she experiences. 

This model has been used to study the local provision of public goods in \cite{naghizadeh15, bramoulle14, bramoulle07}. In the context of security (when viewed as a public good), it is a generalization of the total effort model used in the seminal work of Varian \cite{varian04}, and is similar to the effective investment model of \cite{walrand11}  and the linear influence network game of \cite{miura08}. 

\subsection{Characterizing effort outcomes}\label{sec:lcp-form}
We now consider the problem of finding the efforts at two outcomes of public good provision games: the Nash equilibria and Pareto efficient effort profiles. A Nash equilibrium is an effort profile at which no agent has an incentive to unilaterally deviate from her strategy given other agents' efforts. This is an effort profile that emerges at the status quo as a result of strategic agents' interactions. A Pareto efficient outcome is an effort profile at which it is not possible to increase any agent's utility without making at least one other agent worse off as a result. It is therefore an indication of the profile's efficiency relative to other possible outcomes. These profiles can be attained through negotiation among agents, or following the introduction of appropriate incentives such as monetary taxes/rewards. 


\subsubsection{Nash equilibria}
we start with the Nash equilibria of the public good provision games.\footnote{We consider pure Nash equilibria of the game. Given the strict concavity of the payoffs in \eqref{eq:weff}, playing the average of a set of effort levels leads to a higher payoff than a mixed strategy over that set. As a result, there is no mixed strategy Nash equilibrium for our games.} A Nash equilibrium is a fixed point of the best-reply map. Formally, let $f_i(\mathbf{x}_{-i}; \mathbf{G})$ be the best reply of agent $i$; i.e., the effort that maximizes $i$'s payoff given other agents' profile of efforts $\mathbf{x}_{-i}$ and the dependence matrix $\mathbf{G}$.  For agents with utility \eqref{eq:weff}, this best reply is given by: 
\begin{align}
f_i(\mathbf{x}_{-i}; \mathbf{G})= \max\{0, \bar{q}_i - \sum_{j\in \mathcal{N}_i} g_{ij}x_j\}~,
\label{eq:LBR}
\end{align}
where $\bar{q}_i$ is the effort level at which $b_i'(\bar{q}_i)=c_i$. In other words, $\bar{q}_i$ is the aggregate effort at which $i$'s marginal utility equals her marginal cost.\footnote{\label{f:quad}It is worth mentioning that the best-response mapping of games with linear quadratic payoffs is also of the form \eqref{eq:LBR}. Formally, in a game with linear quadratic payoffs, the utility of agent $i$ is given by \cite{ballester06}: 
\begin{align*}
u_i(\mathbf{x}; \mathbf{G}) = \bar{q}_i x_i - \frac12x_i^2 - \sum_{j\neq i} g_{ij}x_ix_j~,
\end{align*} 
where $\bar{q}_i$ is a given constant. The delinquency games of \cite{ballester10a} and a Cournot competition with heterogeneous goods and network collaboration (in which $g_{ij}$ determines the degree of substitutability of $i$'s good with $j$'s output) are special cases of games with linear-quadratic payoffs; see \cite{bramoulle14,ballester10} for examples. 
All our results regarding Nash equilibria apply to these (as well as other games with linear best-replies of the form \eqref{eq:LBR}) as well.}

For each effort level $x_i$, define a corresponding complementary variable $w_i$.  Then, finding a fixed point of the mapping \eqref{eq:LBR} is equivalent to finding a solution to the following problem: 
\begin{align}
\mathbf{w} - (\mathbf{I} + \mathbf{G})\mathbf{x} = -\mathbf{\bar{q}}~,\notag\\
\mathbf{w}\succeq \mathbf{0}~, ~~ \mathbf{x}\succeq \mathbf{0}~,\notag\\
\mathbf{w}^T\mathbf{x} = 0~.
\label{eq:lcp-ne}
\end{align}
where $\mathbf{\bar{q}}:=\{\bar{q}_1, \ldots, \bar{q}_N\}$, and $\mathbf{I}$ is the $N\times N$ identity matrix. The optimization problem in \eqref{eq:lcp-ne} is an instance of \emph{linear complementarity problems} (LCPs). 


The Linear Complementarity Problem (LCP) refers to a family of problems which arise in solving linear programming and quadratic programming problems, as well as in finding Nash equilibria of bimatrix (two-player non-zero sum) games \cite{cottle68}. For example, the necessary first order optimality (KKT) conditions of a quadratic programming problem constitute an LCP. In addition to these direct connections, LCPs have found applications in the study of market equilibrium, computing Brouwer and Kakutani fixed points, and developing efficient algorithms for solving nonlinear programming problems \cite{murty88}. 

Formally, an LCP $(\mathbf{M}, \mathbf{q})$ is the problem of finding vectors $\mathbf{x}\in \mathbb{R}^n$ and $\mathbf{w}\in \mathbb{R}^n$ satisfying: 
\begin{align}
\mathbf{w} - \mathbf{M}\mathbf{x} = \mathbf{{q}}~,\notag\\
\mathbf{w}\succeq \mathbf{0}~, ~~ \mathbf{x}\succeq \mathbf{0}~,\notag\\
\mathbf{w}^T\mathbf{x} = 0~.
\label{eq:lcp-def}
\end{align}
An LCP is therefore fully determined by an $n\times n$ square matrix $\mathbf{M}$ and a constant right-hand vector $\mathbf{q}\in \mathbb{R}^n$. 

Comparing \eqref{eq:lcp-def} with \eqref{eq:lcp-ne}, we observe that finding the Nash equilibria for the public good provision game is equivalent to solving the LCPs $(\mathbf{(I+G)}, -\mathbf{\bar{q}})$. In Section \ref{sec:e&u}, we will identify conditions on the dependence matrix $\mathbf{G}$ such that solutions to \eqref{eq:lcp-ne} exist and are unique, \emph{for all} right-hand vectors $\mathbf{q}$. In other words, we are interested in the structural properties of the interaction network that guarantee the existence and uniqueness of Nash equilibria, for any payoffs of the form \eqref{eq:weff}, irrespective of the realization of benefit functions or marginal costs of effort. 

\emph{Remark (on the sign of $\mathbf{q}$):} We note that the right-hand vector entries $q_i$ in \eqref{eq:lcp-def} can be either positive, negative, or zero. In particular, for $\mathbf{q}\succ\mathbf{0}$ in \eqref{eq:lcp-def}, the LCP always has the solution $\mathbf{w}=\mathbf{q}$ and $\mathbf{x}=\mathbf{0}$. In the case of Nash equilibria with LCP $(\mathbf{I+G}, -\mathbf{\bar{q}})$, $\mathbf{\bar{q}} \prec \mathbf{0}$ implies that the zero effort profile $\mathbf{x}=\mathbf{0}$ is always a possible Nash equilibrium. This observation can be intuitively explained as follows. Recall that $q_i$ indicates the effort level at which agent $i$'s marginal utility equals her marginal cost. A negative $q_i$ therefore indicates that exerting effort is not cost-efficient for agent $i$. Hence, a zero effort equilibrium is indeed to be expected.

\subsubsection{Pareto efficient effort profiles} we also consider Pareto efficient effort profiles of the public good provision game. {Formally, we consider the solutions to the following problem: 
\begin{align*}
\max_{\mathbf{x}\succeq \mathbf{0}} \sum_i \lambda_i u_i(\mathbf{x})
\end{align*}
where $\boldsymbol{\lambda}:=\{\lambda_1, \cdots, \lambda_N\}$ is a vector of non-negative weights. By \cite[Proposition 16.E.2]{mas-micro}, for the strictly concave utility functions $u_i(\cdot)$ given by \eqref{eq:weff}, the set of solutions to this linear welfare maximization problem, as $\boldsymbol{\lambda}$ ranges over the set of all strictly positive weight vector, leads to the Pareto optimal effort profiles.} It is worth noting that solving for the Pareto efficient profile with unit vector of weights $\boldsymbol{\lambda}=\mathbf{1}$ in \eqref{eq:pe-lambda} will lead to the socially optimal profile of efforts $\mathbf{x}^*=\arg\max_{\mathbf{x}\succeq \mathbf{0}} \sum_i u_i(\mathbf{x})$. 

We now proceed to characterizing these profiles. Consider the Pareto efficient effort profile $\mathbf{x}^{\boldsymbol{\lambda}}$ corresponding to the strictly positive weight vector $\boldsymbol{\lambda}$.  That is,
\begin{align}
\mathbf{x}^{\boldsymbol{\lambda}} = \arg \max_{\mathbf{x}\succeq \mathbf{0}} \sum_k \lambda_k u_k(\mathbf{x})~.
\label{eq:pe-lambda}
\end{align}
The first order condition on \eqref{eq:pe-lambda} with respect to $x_i$ implies that at the Pareto efficient solution, the following should hold: 
\begin{align}
  b_i'(x^{\boldsymbol{\lambda}}_i+ \sum_{j\in \mathcal{N}_i} g_{ij}x^{\boldsymbol{\lambda}}_j) + \sum_{k, \text{ s.t. } i\in \mathcal{N}_k} \frac{\lambda_k}{\lambda_i} g_{ki} b_k'(x^{\boldsymbol{\lambda}}_k + \sum_{j\in \mathcal{N}_k} g_{kj}x^{\boldsymbol{\lambda}}_j)  = c_i - z_i,~  \forall i~.
  \label{eq:pe-foc}
\end{align}
Here, $z_i$ is a complementary variable corresponding to the effort level $x^{\boldsymbol{\lambda}}_i$. 

Consider an \emph{interior} Pareto efficient outcome in which all agents exert non-zero effort; i.e., $\mathbf{z}=\mathbf{0}$. We will study graph-theoretical characterizations of these outcomes, as well as interior Nash equilibria, in Section \ref{sec:cent}. Define $\mathbf{q}^{\boldsymbol{\lambda}}$ as the effort levels satisfying:
\begin{align}
b_i'(q^{\boldsymbol{\lambda}}_i) + \sum_{k, \text{ s.t. } i\in \mathcal{N}_k} \frac{\lambda_k}{\lambda_i}  g_{ki} b_k'(q^{\boldsymbol{\lambda}}_k)  = c_i~, \forall i.
\label{eq:pe-interior}
\end{align}
Intuitively, $\mathbf{q}^{\boldsymbol{\lambda}}$ is the vector of efforts at which the marginal \emph{social} benefits equal the marginal (social) costs of effort. When $\mathbf{I}+\mathbf{G}$ is invertible, we can find the following alternative expression for $q_i^{\boldsymbol{\lambda}}$ by solving the system of equations in \eqref{eq:pe-interior}: 
\[b_i'(q_i^{\boldsymbol{\lambda}})=\left((\mathbf{I} + \boldsymbol{\Lambda}^{-1}\mathbf{G}^T\boldsymbol{\Lambda})^{-1} \mathbf{c} \right)_i~.\]
Here, $q_i^{\boldsymbol{\lambda}}$ can be interpreted as the aggregate effort level at which agent $i$'s marginal benefit equals her \emph{modified} marginal cost. The modification depends on the graph structure, as well as the weights $\boldsymbol{\lambda}$. We will elaborate further in Section \ref{sec:cent-eff}. 

Finding such interior Pareto efficient effort profile $\mathbf{x}^{\boldsymbol{\lambda}}$ is equivalent to finding a solution with $\mathbf{w}=\mathbf{0}$ to the following problem: 
\begin{align}
\mathbf{w} - (\mathbf{I} + \mathbf{G})\mathbf{x} = -\mathbf{{q}^{\boldsymbol{\lambda}}}~,\notag\\
\mathbf{w}\succeq \mathbf{0}~, ~~ \mathbf{x}\succeq \mathbf{0}~,\notag\\
\mathbf{w}^T\mathbf{x} = 0~.
\label{eq:lcp-pe-interior}
\end{align}
In other words, finding interior Pareto efficient outcomes is equivalent to finding solutions to the LCP $((\mathbf{I}+\mathbf{G}), -\mathbf{{q}^{\boldsymbol{\lambda}}})$ with $\mathbf{w}=\mathbf{0}$. We study conditions under which such solutions exist in Section \ref{sec:interior-exist}. 


\section{Existence and Uniqueness of Nash Equilibria} \label{sec:e&u}

In this section, we study conditions under which Nash equilibria of public good provision games exist, and in particular, conditions under which these profiles are unique.  We contrast our result with those in the existing literature, and show how existing conditions can be recovered as corollaries of our main theorem. 

\subsection{Existence and uniqueness} \label{sec:exist-unique}

Using the LCP formulation of the problem of finding the Nash equilibria in \eqref{eq:lcp-ne}, we identify conditions for the existence and uniqueness of this equilibrium. We begin with the following definition. 

\begin{definition}
A square matrix $\mathbf{M}$ is a \emph{P-matrix} if the determinants of all its principal minors (i.e., the square submatrix obtained from $\mathbf{M}$ by removing a set of rows and their corresponding columns) are strictly positive. 
\end{definition} 

The following theorem provides a necessary and sufficient condition under which the Nash equilibrium exists and is unique. 

\vspace{0.16in}
\begin{theorem}[Uniqueness]\label{t:uniqueness}
The public good provision game has a unique Nash equilibrium if and only if $\mathbf{I}+\mathbf{G}$ is a P-matrix. 
\end{theorem}
\vspace{0.16in} 

The proof follows from results on the uniqueness of solutions of LCPs; see e.g.,  \cite[Theorem 4.2]{murty72}. We illustrate Theorem \ref{t:uniqueness} through an example. 

\begin{example}\label{ex:unique}
Consider a network of two nodes. We study the Nash equilibria of a public good provision game with payoffs:
\begin{align*}
u_i (\mathbf{x}; \mathbf{G}) = 1-\exp(-x_i-g_{ij}x_j) - \frac{1}{e}x_i, \text{for } i\in \{1,2\}, j\neq i~.
\end{align*}
Note that $\mathbf{I} + \mathbf{G}$ is a P-matrix if and only if $g_{12}g_{21}<1$. 

\emph{(i)} First, let $g_{12}=g_{21}=\frac12$. Then, by Theorem \ref{t:uniqueness}, this game should have a unique Nash equilibrium. Indeed, this unique equilibrium is given by $x_1=x_2=\frac23$. 

\emph{(ii)} Next, consider $g_{12}=g_{21}=2$. Then $\mathbf{I}+ \mathbf{G}$ is not a P-matrix, and the game need not have a unique Nash equilibrium. For the given payoffs, there are three possible Nash equilibria:  $(x_1, x_2) = (0,1)$, $(x_1, x_2) = (1,0)$, and $(x_1, x_2) = (\frac13, \frac13)$. 

\emph{(iii)} Finally, let $g_{12}=g_{21}=-2$. Again, $\mathbf{I}+ \mathbf{G}$ is not a P-matrix, and hence by Theorem \ref{t:uniqueness}, the corresponding game need not have a unique equilibrium. In fact, under the assumed payoff functions, the game will have no Nash equilibrium. 
\end{example}

\vspace{0.1in}

We now turn to the more general question of \emph{existence} of Nash equilibria. We are interested in weaker conditions than those of Theorem \ref{t:uniqueness} that guarantee at least one Nash equilibrium exists. Unlike uniqueness, there is no simple characterization of matrices $\mathbf{M}$ for which an LCP $(\mathbf{M}, \mathbf{q})$ has a solution. Nevertheless, we can identify existence results on two particular subclasses of games, namely games of strategic substitutes and games of strategic complements. Recall that for a game of strategic substitutes (complements), $g_{ij}\geq 0 ~(g_{ij}\leq 0), \forall i, j\neq i$. 

\vspace{0.16in}
\begin{theorem}[Existence in games with strategic substitutes]\label{t:existence-sub}
A public good provision game with strategic substitutes always has at least one Nash equilibrium. 
\end{theorem}
\vspace{0.1in}
\begin{IEEEproof}
By  \cite[Theorem 5.2]{murty72}, for a given non-negative matrix $\mathbf{M}$, the corresponding LCP $(\mathbf{M}, \mathbf{q})$ has a solution for all $\mathbf{q}$ if and only if $m_{ii}>0$. For a game with substitutes, $\mathbf{I} + \mathbf{G}$ is a non-negative matrix, and the diagonal entries are all 1. Therefore, for LCP \eqref{eq:lcp-ne},   a solution (Nash equilibrium) always exists. 
\end{IEEEproof}
\vspace{0.16in}

We next consider the existence of Nash equilibria in games where agents' efforts are complements to their neighbors'. Let $\rho(\mathbf{G}):=\max\{|\lambda| \text{ s.t. } \mathbf{G}\mathbf{v} = \lambda\mathbf{v}\}$ denote the spectral radius of $\mathbf{G}$. Also, define the following classes of matrices. 

\begin{definition}[Z-matrix, L-matrix, S-matrix]
\begin{itemize}
\item A square matrix $\mathbf{M}$ is a \emph{Z-matrix} if $m_{ij}\leq 0, \forall i, j\neq i$. 
\item A square matrix $\mathbf{M}$  is an \emph{L-matrix} if it is Z-matrix and $m_{ii}>0, \forall i$. 
\item A matrix $\mathbf{M}$ is an \emph{S-matrix} if there exists $\mathbf{x}\succ \mathbf{0}$ such that $\mathbf{M}\mathbf{x}\succ \mathbf{0}$. 
\end{itemize}
\end{definition}

\vspace{0.16in}
\begin{theorem}[Existence in games with strategic complements]\label{t:existence-comp}
For a public good provision game with strategic complements, if a Nash equilibrium exists for all $\mathbf{\bar{q}}$, i.e., for all payoff realizations, then it is unique. Specifically, the game has a Nash equilibrium if and only if $\rho(\mathbf{G})<1$. 
\end{theorem}
\vspace{0.1in}
\begin{IEEEproof}
First, note that for this game, $\mathbf{I}+\mathbf{G}$ is an L-matrix. For an LCP $(\mathbf{M}, \mathbf{q})$, if $\mathbf{M}$ is an L-matrix, the LCP has at least one solution for all $\mathbf{q}$ if and only if $\mathbf{M}$ is an S-matrix; see \cite[p. 282]{murty88}. Therefore, the LCP \eqref{eq:lcp-ne} has a solution if and only if $\mathbf{I}+\mathbf{G}$ is an S-matrix. A Z-matrix is an S-matrix if and only if it is a P-matrix \cite{pang79}. Therefore, the condition for existence and uniqueness in games with complements are the same. In other words, if a Nash equilibrium is guaranteed to exist, it is also unique. 

Also, for a Z-matrix $\mathbf{G}$, $\mathbf{I}+\mathbf{G}$ is an S-matrix if and only $\rho(\mathbf{G})<1$ \cite{berman94}. Therefore, a solution exists and is unique if and only if $\rho(\mathbf{G})<1$.\footnote{The statement of Theorem \ref{t:existence-comp} is similar to Theorem 1 in \cite{corbo07}, which also uses an LCP formulation in the study of Nash equilibria on unweighted and undirected networks where agents have linear quadratic payoffs. This can be explained by observing that both games have best replies of the form \eqref{eq:LBR}, and hence have similar conclusions; c.f. footnote \ref{f:quad}.} 
\end{IEEEproof}
\vspace{0.1in}

It is worth noting the difference between Theorems \ref{t:existence-sub} and \ref{t:existence-comp} and a previous result on the existence of Nash equilibria in concave n-person games. Rosen \cite{rosen65} shows that for an n-person game, if agents' payoffs are concave in their own effort, and agents' strategies are limited to a convex, closed, and bounded set, then the corresponding n-person game always has a Nash equilibrium \cite[Theorem 1]{rosen65}. The latter assumption does not hold in the current model, as we allow an unbounded effort space $x_i\in\mathbb{R}_{\geq 0}$.  

However, similar to \cite{rosen65}, Theorem \ref{t:existence-sub} concludes that for games of strategic substitutes, a Nash equilibrium always exists.  In this case, each agent's strategy space can be effectively bounded by $q_i$, where $b_i'(q_i)=c_i$, i.e., agent $i$ may exert effort lower than $q_i$ (due to positive externalities from her neighbors), but will never exert an effort higher than $q_i$, as her marginal cost to do so will be higher than her marginal benefit.  Thus in this case the existence result given by Theorem \ref{t:existence-sub} is equivalent to that given in \cite{rosen65}, though arrived at using a different methodology. 

For games of strategic complements on the other hand, a similar upper bound on agents' strategies does not exist. Specifically, when an agent $i$'s neighbors increase their efforts, she will experience a negative externality, and will therefore increase her own level of effort to compensate for the lost benefit. As a result, agents' efforts can grow unbounded, and an equilibrium may not exist; the sufficient and necessary condition given in Theorem \ref{t:existence-comp} thus goes beyond that considered in \cite{rosen65}. 
If the strategy spaces were bounded in this scenario, then agents would exert the upper bound effort, leading to the existence result of \cite{rosen65}.

\subsection{Comparison with existing results} \label{sec:compare}

We now show how existing results in \cite{bramoulle14,ballester10,naghizadeh15,miura08} on the uniqueness of the Nash equilibrium of public good provision games can be recovered as corollaries of Theorem \ref{t:uniqueness}. These comparisons also illustrate that some well-known matrices, namely, symmetric positive definite, strongly diagonally dominant, and (a subclass of) Z-matrices, belong to the family of P-matrices. 

We begin with the uniqueness result of \cite{bramoulle14} on networks of symmetric relations. We note that \cite{bramoulle14} only states the sufficient condition; we also show the necessary condition in the following corollary using Theorem \ref{t:uniqueness}.  
\vspace{0.16in}
 
\begin{corollary}[Uniqueness on symmetric networks \cite{bramoulle14}]\label{c:kranton}
Consider a network with a symmetric dependence matrix $\mathbf{G}$. Then, if and only if $|\lambda_{min}(\mathbf{G})|<1$, the Nash equilibrium is unique.  
\end{corollary}
\vspace{0.1in}
\begin{IEEEproof}
By \cite[Theorem 1.9]{murty72} a square symmetric matrix is a P-matrix if and only if it is positive definite. Therefore, by Theorem \ref{t:uniqueness}, the Nash equilibrium is unique if and only if $\mathbf{I}+\mathbf{G}$ is positive definite, which occurs if and only if $|\lambda_{min}(\mathbf{G})|<1$.
\end{IEEEproof}
\vspace{0.1in}

The results of \cite{bramoulle14} are the first to show the importance of the lowest eigenvalue in determining outcomes of strategic interactions on networks, leading to several interesting insights on equilibria stability and network structure; we refer the interested reader to \cite{bramoulle14} for details. 

It is also worth noting that Theorem \ref{t:uniqueness} generalizes \cite{bramoulle14} on both symmetric and asymmetric matrices: 

\emph{(i) Symmetric matrices:} \cite{bramoulle14} uses the theory of potential games to show that a positive definite $\mathbf{I}+\mathbf{G}$ is a sufficient condition for uniqueness of the Nash equilibrium. Our result shows that this condition is necessary as well. 

\emph{(ii) Asymmetric matrices:} For directed, asymmetric graphs, the results of \cite{bramoulle14} apply if $|\lambda_{min}(\frac{\mathbf{G}+\mathbf{G}^T}{2})|<1$; i.e., if $\mathbf{I} + \frac{\mathbf{G}+\mathbf{G}^T}{2}$ is positive definite. This is equivalent to $\mathbf{I}+\mathbf{G}$ being positive definite \cite[Result 1.9]{murty72}. In contrast, Theorem \ref{t:uniqueness} only requires that $\mathbf{I}+\mathbf{G}$ be a P-matrix, providing a more general (weaker) sufficient condition (as well as a necessary condition).  
This is because there exist (asymmetric) P-matrices that are not positive definite \cite[Theorem 1.10]{murty72}. Hence, positive definite matrices are in general a subset of P-matrices. 

\vspace{0.05in}
We next show that the result of \cite{ballester10} can also be recovered as a corollary of Theorem \ref{t:uniqueness}. 
\vspace{0.1in}
\begin{corollary}[Uniqueness on networks with hidden complementarities \cite{ballester10}]\label{c:ballester}
Let $\mathbf{T}$ be a Z-matrix such that $\mathbf{T}(\mathbf{I}+\mathbf{G})$ is both a Z-matrix and an S-matrix. Then, the Nash equilibrium is unique if and only if $\mathbf{I}+\mathbf{G}$ is an S-matrix. In particular, if $\mathbf{G}$ is a Z-matrix (i.e., a game with complementarities), the equilibrium is unique if and only if $\rho(\mathbf{G})<1$. 
\end{corollary}
\vspace{0.1in}
\begin{IEEEproof}
By Theorem \ref{t:uniqueness}, we know that the Nash equilibrium is unique if and only if $\mathbf{I}+\mathbf{G}$ is a P-matrix. On the other hand, a matrix $\mathbf{I}+\mathbf{G}$ satisfying the conditions of the corollary is a \emph{hidden Z-matrix} \cite{pang79}. By \cite[Theorem 1]{pang79}, a hidden Z-matrix is a P-matrix if and only if it is an S-matrix. Therefore, the Nash equilibrium is unique if and only if $\mathbf{I}+\mathbf{G}$ is an S-matrix. Finally, when $\mathbf{G}$ is a Z-matrix, $\mathbf{I}+\mathbf{G}$ is an S-matrix if and only $\rho(\mathbf{G})<1$ \cite{berman94}. 
\end{IEEEproof}
\vspace{0.1in}

We also prove an alternative expression for Corollary \ref{c:ballester}. 
\begin{corollary}\label{c:ballester-alt}
If $\mathbf{G}$ is a Z-matrix, a unique Nash equilibrium exists if and only if $|\lambda_{min}(\mathbf{G})|<1$. 
\end{corollary}
\begin{IEEEproof}
For a Z-matrix $\mathbf{G}$, $-\mathbf{G}$ is a non-negative matrix. Then, by the Perron-Frobenius theorem, $-\mathbf{G}$ has a positive eigenvalue equal to its spectral radius, $\lambda_{max}(-\mathbf{G})=\rho(-\mathbf{G})$. Noting that $\rho(-\mathbf{G})= \rho(\mathbf{G})$ and $\lambda_{max}(-\mathbf{G})=-\lambda_{min}(\mathbf{G})$, we conclude that for Z-matrices, $\rho(\mathbf{G})<1$ if and only if $|\lambda_{min}(\mathbf{G})|<1$. 
\end{IEEEproof}

Comparing the above with Corollary \ref{c:kranton}, we conclude that the lowest eigenvalue of the dependence matrix has the key role in determining sufficient (and necessary) conditions for the uniqueness of Nash equilibria in (asymmetric) networks with complementarities (in addition to the symmetric networks and some subclasses of directed networks shown in \cite{bramoulle14}). 

\vspace{0.05in}
Finally, we show that the result of \cite{naghizadeh15, miura08} can also be recovered as a corollary of Theorem \ref{t:uniqueness}. 
\vspace{0.1in}
\begin{corollary}[Uniqueness on strictly diagonally dominant networks \cite{naghizadeh15, miura08}]\label{c:miura-ko}
If $\mathbf{I}+\mathbf{G}$ is strictly diagonally dominant; i.e., $\sum_{i}|g_{ij}|<1,\forall i$, there is a unique Nash equilibrium. 
\end{corollary}
\vspace{0.1in}
\begin{IEEEproof}
We prove the theorem by showing that if $\mathbf{I}+\mathbf{G}$ is strictly diagonally dominant, then it is a P-matrix. This is because by the Gershgorin circle theorem, for a strictly diagonally dominant matrix with positive diagonal elements, all real eigenvalues are positive. Following a similar argument, all real eigenvalues of all sub-matrices of $\mathbf{I}+\mathbf{G}$ are also positive. Since the determinant of a matrix is the product of its eigenvalues, and as for real matrices, the complex eigenvalues appear in pairs with their conjugate eigenvalues, it follows that $\mathbf{I}+\mathbf{G}$, as well as all its square sub-matrices, have positive determinants. Therefore, $\mathbf{I}+\mathbf{G}$ is a P-matrix. The uniqueness then follows from Theorem \ref{t:uniqueness}.
 \end{IEEEproof}
\vspace{0.1in}

\section{Efforts as Node Centralities} \label{sec:cent}

In this section, we focus on \emph{interior} effort profiles of the public good provision game; that is, outcomes in which all agents exert strictly positive efforts. We establish a connection between agents' actions at interior Nash equilibria, as well as interior Pareto efficient outcomes, and agents' centralities in the dependence network. Using this connection, we can identify the effects of dependencies (outgoing edges in $\mathbf{G}$) and influences (incoming edges in $\mathbf{G}$), as well as walks of different length, on the efforts exerted by agents. 

\subsection{Existence of interior effort profiles} \label{sec:interior-exist}

We first identify conditions under which a game with payoffs \eqref{eq:weff} has interior Nash equilibria and Pareto efficient effort profiles. We begin with a definition. 
\begin{definition}[Positive cone]\label{def:pos}
The positive cone (or positive linear span) of a set of vectors $\mathbf{v}=\{v_1, v_2, \ldots, v_n\}$ is given by $pos(\mathbf{v}):=\{\sum_i \alpha_iv_i | ~ \alpha_i\geq 0, \forall i\}$. 
\end{definition}

For a Nash equilibrium (or a Pareto efficient effort profile) to be interior, the corresponding LCP \eqref{eq:lcp-ne} (or \eqref{eq:lcp-pe-interior}) should have a solution with $\mathbf{x}\succeq \mathbf{0}, \mathbf{w} = \mathbf{0}$.\footnote{With a slight abuse of terminology, we consider solutions with $x_i=0, w_i=0$ to be interior as well.} 

\vspace{0.16in}
\begin{theorem}[Existence of interior effort profiles]\label{t:interior}
A public good provision game with payoffs \eqref{eq:weff} has an interior Nash equilibrium (or Pareto efficient effort profile) if and only if the corresponding $\mathbf{\bar{q}}$ (or $\mathbf{q}^{\boldsymbol{\lambda}}$) is in the positive cone generated by the columns of $\mathbf{I}+\mathbf{G}$. 
\end{theorem}
\vspace{0.1in}
\begin{IEEEproof}
Solving the LCP \eqref{eq:lcp-ne} for interior solutions is equivalent to finding a solution to: 
\[(\mathbf{I}+\mathbf{G})\mathbf{x} = \mathbf{\bar{q}}, ~~ \mathbf{x}\succeq \mathbf{0}~.\]
The theorem then follows from Definition \ref{def:pos}. The same argument applies to finding interior Pareto efficient profiles using \eqref{eq:lcp-pe-interior}.  
It is also worth mentioning that given $\mathbf{G}$, non-interior solutions will necessarily exist for some $\mathbf{q}\in\mathbb{R}^n$, as we need at least $n+1$ vectors to positively span $\mathbb{R}^n$ \cite[Theorem 3.8]{davis54}. In other words, as expected, there is no network structure for which solutions are guaranteed to be interior. 
\end{IEEEproof}
\vspace{0.1in}

We now proceed to establishing a connection between interior effort profiles (when they exist) and agents' centralities in their interaction network, starting with an overview of centrality measures.

\subsection{Alpha-centrality: an overview}\label{sec:centrality}

Centrality measures have been used extensively in the graph theory and network analysis literatures as indicators of importance of nodes in their interaction network. Some of these measures (e.g., degree centrality) take into account the number of connections of a node in determining her centrality. In contrast, another class of measures (e.g. eigenvalue centrality) account for the importance of the connections as well, such that a node's centrality is (recursively) related to those of her neighbors. \emph{Alpha-centrality}, considered herein, belongs to the latter family. This measure was introduced by Bonacich and Lloyd in \cite{bonacich01}, mainly as an extension of eigenvalue centrality that is applicable to networks of asymmetric relations.

Formally, denote the centrality of node $i$ by $x_i$. Let $\mathbf{G}$ be the adjacency matrix of a network, where $g_{ij}$ determines the dependence of node $i$ on node $j$. 
Then, the eigenvalue centrality of nodes will be proportional to $\mathbf{G}\mathbf{x}$. Alpha-centrality generalizes this measure by allowing the nodes to additionally experience an exogenous source of centrality $\mathbf{e}$, such that: 
\[\mathbf{x} = \alpha \mathbf{G} \mathbf{x} + \mathbf{e}~.\]
Here, $\alpha$ is a constant that determines a tradeoff between the endogenous (eigenvalue) and exogenous centrality factors. The nodes' alpha-centralities are therefore given by: 
\begin{align}
c_\text{alpha}(\mathbf{G}, \alpha, \mathbf{e}) = (\mathbf{I}-\alpha \mathbf{G})^{-1}\mathbf{e}~.
\label{eq:alpha-cent}
\end{align}

\paragraph*{On the interpretation of $\alpha$} as mentioned above, $\alpha$ determines the tradeoff between the endogenous and exogenous sources of centrality. We will now illustrate that powers of $\alpha$ also appear as weights of walks of different length in determining nodes' centralities.  

We do so by noting the connection between alpha-centrality and the measure proposed by Katz \cite{katz53}. Katz centrality defines a weighted sum of powers of the adjacency matrix $\mathbf{G}$ as an indicator of nodes' importance; intuitively, longer walks are weighed differently (and often less favorably) in determining nodes' centralities. Formally, Katz' measure is given by:
\[c_\text{katz}(\mathbf{G}, \alpha) = (\sum_{i=1}^\infty \alpha^i \mathbf{G}^i)\mathbf{1}~,\] 
where $\alpha$ is an attenuation factor. In particular, if $\alpha<\frac{1}{|\lambda_{max}(\mathbf{G})|}$, the infinite sum converges, so that: 
\begin{align}
(\sum_{i=1}^\infty \alpha^i \mathbf{G}^{i})\mathbf{e} = (-I+(I-\alpha \mathbf{G})^{-1})\mathbf{e}~.
\label{eq:katz-cents}
\end{align}
Comparing \eqref{eq:alpha-cent} and \eqref{eq:katz-cents}, we conclude that the parameter $\alpha$ of alpha-centrality can be similarly interpreted as a weight assigned to the walks of different length in determining the effect of endogenous centralities on the overall centrality of a node.
\footnote{\label{f:bon-alpha}Alpha centrality is also similar to the measure introduced earlier by Bonacich in his seminal work \cite{bonacich87}. Formally, Bonacich's centrality is defined as $c_\text{bonacich}(R, \beta, \alpha) =  \beta(I - \alpha R)^{-1}R \mathbf{1}$. Here, $R$ is a symmetric matrix of relationships, with main diagonal elements equal to zero. The parameter $\beta$ only affects the length of the final measures, and has no network interpretation. The parameter $\alpha$ on the other hand can be positive or negative, and determines the extent and direction of influences. On symmetric matrices, Katz' measure is essentially equivalent to Bonacich centrality; in fact, $c_\text{katz}(R, \alpha) = \sum_{i=1}^\infty \alpha^i R^i \mathbf{1} = \alpha c_\text{bonacich}(R, \alpha, 1)$.  
To summarize, taking the three measures on a symmetric matrix $A$, and setting $\mathbf{e}=\mathbf{1}$ for the alpha-centralities, we have: 
\[c_\text{alpha}(A, \alpha, \mathbf{1}) = 1 + \alpha c_\text{bonacich}(A,\alpha, 1) = 1 + c_\text{katz}(A,\alpha)~.\]
Therefore, in essence, alpha-centrality generalizes Bonacich and Katz centralities, allowing for vectors of exogenous status $\mathbf{e}$. Using the above equivalence, we can show that our characterization of Nash equilibrium based on alpha-centralities in Theorem \ref{t:cent-eff}, and the Nash-Bonacich linkage established in \cite{ballester06} are equivalent (see also footnote \ref{f:quad}).}

\subsection{A centrality-effort connection}\label{sec:cent-eff}
We now establish the connection between agents' efforts at interior profiles, and their alpha-centralities in the interaction network. 

\vspace{0.16in}
\begin{theorem}[Centrality-effort connection]\label{t:cent-eff} 

(i) Consider an interior Nash equilibrium $\mathbf{x}^*$. Then, 
\[\mathbf{x}^* = c_\text{alpha} \left(\mathbf{G}, -1, \mathbf{\bar{q}}\right)~,\]
where $\mathbf{\bar{q}}$ is such that $b'_i(\bar{q}_i)=c_i$. 

(ii) Consider an interior Pareto efficient effort profile $\mathbf{x}^{\boldsymbol{\lambda}}$. Then, 
\[\mathbf{x}^{\boldsymbol{\lambda}} = c_\text{alpha} \left(\mathbf{G}, -1, \mathbf{q}^{\boldsymbol{\lambda}}\right)~,\]
where $\mathbf{q}^{\boldsymbol{\lambda}}$ is such that $b'_i(q^{\boldsymbol{\lambda}}_i)=c_{\text{alpha}, i}(\boldsymbol{\Lambda}^{-1}\mathbf{G}^T\boldsymbol{\Lambda}, -1, \mathbf{c})$. 
\end{theorem}
\vspace{0.1in}
\begin{IEEEproof}
(i) An interior Nash equilibrium is a solution to LCP \eqref{eq:lcp-ne} with $\mathbf{w}=\mathbf{0}$; i.e.,
\begin{align*}
(\mathbf{I}+\mathbf{G})\mathbf{x} = \mathbf{\bar{q}}, ~~ \mathbf{x}\succeq \mathbf{0}~.
\end{align*}
Therefore, when such solution exists, $\mathbf{x}^* = (\mathbf{I}+\mathbf{G})^{-1}\mathbf{\bar{q}}$. Comparing this expression with \eqref{eq:alpha-cent} establishes the connection. 

(ii) An interior Pareto efficient profile with weights $\boldsymbol{\lambda}$ is a solution to LCP \eqref{eq:lcp-pe-interior} with $\mathbf{w}=\mathbf{0}$; i.e.,
\begin{align*}
(\mathbf{I}+\mathbf{G})\mathbf{x} = \mathbf{q}^{\boldsymbol{\lambda}}, ~~ \mathbf{x}\succeq \mathbf{0}~.
\end{align*}
Therefore, when such solution exists, $\mathbf{x}^{\boldsymbol{\lambda}} = (\mathbf{I}+\mathbf{G})^{-1}\mathbf{q}^{\boldsymbol{\lambda}}$. Also, by definition, we know that $\mathbf{q}^{\boldsymbol{\lambda}}$ satisfies $b_i'(q_i^{\boldsymbol{\lambda}})=\left((\mathbf{I} + \boldsymbol{\Lambda}^{-1}\mathbf{G}^T\boldsymbol{\Lambda})^{-1} \mathbf{c} \right)_i.$ Comparing these expressions with \eqref{eq:alpha-cent} establishes the connection. 
\end{IEEEproof}

\vspace{0.1in}

The connection established in Theorem \ref{t:cent-eff} leads to several interesting insights. Recall that an entry $g_{ij}\neq 0$ in $\mathbf{G}$ indicates that agent $i$'s payoff depends on agent $j$'s action; we therefore refer to $\mathbf{G}$ as the \emph{dependence} matrix. On the other hand, an entry $g_{ji}\neq 0$ in the $\mathbf{G}^T$ indicates that agent $j$'s effort influences agent $i$'s payoff. We will therefore refer to $\mathbf{G}^T$ as the \emph{influence} matrix. 

\vspace{0.1in}
\textbf{Perceived costs at different effort profiles:}  comparing parts (i) and (ii) of Theorem \ref{t:cent-eff}, we observe that the only difference when determining nodes' efforts is in the corresponding vectors of exogenous centralities. These vectors are determined by efforts at which agents' marginal benefits equal their (perceived) marginal costs. At the Nash equilibrium, each agent acts independently and perceives only her own cost of effort, leading to $b'_i(\bar{q}_i)=c_i$. On the other hand, for Pareto efficient solutions to emerge, the cost perceptions are modified according to agents' positions in the network, as well as the importance placed on each agent's welfare, as determined by $\lambda_i$. Consequently, both $\mathbf{G}$ and $\boldsymbol{\lambda}$ play a role in determining agents' perceived marginal costs, leading to $b'_i(q^{\boldsymbol{\lambda}}_i)=c_{\text{alpha}, i}(\boldsymbol{\Lambda}^{-1}\mathbf{G}^T\boldsymbol{\Lambda}, -1, \mathbf{c})$. 

\vspace{0.1in}
\textbf{Effects of dependencies:} consider agents' dependencies (outgoing edges in the network). We observe that by the definition of alpha-centrality \eqref{eq:alpha-cent}, the matrix of dependencies $\mathbf{G}$ shapes the endogenous component of the centrality measure, determining a node's centrality as a function of her neighbors' centrality. Similarly, $\mathbf{G}$ in Theorem \ref{t:cent-eff} indicates that the dependence of an agent on her neighbors (and the efforts they have exerted) will shape her final effort. Note also that this is the case for both Nash equilibria (part (i)) and Pareto efficient efforts (part (ii)): an agent benefits of any neighbor's effort regardless of the solution concept, mechanism, or negotiations through which the effort profile is implemented.

\vspace{0.1in}
\textbf{Effects of influences:} we further observe the effects of agents' influences (incoming edges in the network) on the outcomes of their strategic interactions. The matrix of influences $\mathbf{G}^T$ appears when determining the perceived costs of agents in Pareto efficient solutions. Intuitively, an agent with higher influence on others (as determined by her alpha-centrality in the network of influences) will have a lower perceived marginal cost, hence a higher exogenous centrality (due to concavity of $b_i(\cdot)$), which in turn increases a node's alpha-centrality (i.e, her level of effort/contribution). Note also that the matrix of influences $\mathbf{G}^T$ does not appear in the characterization of the Nash equilibria in Theorem \ref{t:cent-eff}. This is because at a Nash equilibrium, an agent only accounts for her own marginal costs when selecting an effort level.

\vspace{0.1in}
\textbf{Alternating effect -- the role of $\alpha$:} most interestingly, we note that the alpha parameter of all the alpha-centralities in Theorem \ref{t:cent-eff} is $\alpha=-1$. Recall that, as shown in Section \ref{sec:centrality}, $\alpha^k$ is a weight associated with a walk of length $k$ in determining an agent's centrality.\footnote{Given $\alpha=-1$, the condition $\alpha<\frac{1}{|\lambda_{max}(\mathbf{G})|}$ holds for all adjacency matrices $\mathbf{G}$. Therefore, the alpha-centralities can be interpreted as the limit of a weighted sum of powers of the adjacency matrix, and the interpretation of $\alpha$ as a weight on walks of different length in applicable.} Let $i_0, i_1, \ldots, i_k$ be the agents along this walk. Then, for walks of odd length, $\alpha=-1$ induces a sign reversal on the weight  $g_{i_0i_1}g_{i_1i_2}\ldots g_{i_{k-1}i_k}$ of the walk. For walks of even length on the other hand, $\alpha=-1$ leaves the sign on the weight associated with the walk unchanged. 

To better highlight the intuition behind this observation, consider a network of substitutes; i.e., $g_{ij}\geq 0, \forall i,j$. Consider a walk of length one by choosing a neighbor $j$ of $i$. If agent $j$ increases her effort, agent $i$ benefits from the positive externality of $j$'s increased effort, and can in turn reduce her effort. Thus, changes along this walk of odd length negatively affect agent $i$'s effort decision; this is consistent with $(-1)^1g_{ij}<0$. Now, consider a neighbor $k$ of $j$. Therefore, there is a walk of length 2 from $i$ to $k$. By the same argument as above, if $k$ increases her effort, $j$ will decrease her effort in response. To compensate for the lost externality, agent $i$ will now have to increase her own effort. Thus, a change along this walk of even length positively affects agent $i$'s effort decision, which is again consistent with $(-1)^2g_{ij}g_{jk}>0$. The same argument extends to walks of longer lengths. 

Alternatively, consider a network of complementarities; i.e., $g_{ij}\leq 0, \forall i,j\neq i$. Again, consider a walk of length one from $i$ to $j$. If agent $j$ increases her effort, agent $i$'s benefit is reduced, and so she will increase her level of effort in response. Thus, a change along this walk of odd length positively affects agent $i$'s effort decision, which is consistent with $(-1)^1g_{ij}>0$. Now, consider a walk of length 2 from $i$ to $k$ ($j$'s neighbor). By the same argument as above, if $k$ increases her effort, $j$ will increase her effort in response, and so agent $i$ will have to increase her effort as well. Thus, the change along this walk of even length also positively affects agent $i$'s effort decision; this is again consistent with $(-1)^2g_{ij}g_{jk}>0$. The same argument extends to walks of longer lengths.


\subsection{Numerical examples}
We illustrate the centrality-effort connection through some examples.  

\begin{figure}
\centering
\includegraphics{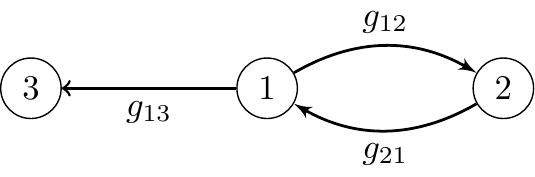}
\\
\includegraphics{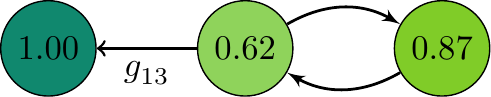}
\hspace{0.1in}
\includegraphics{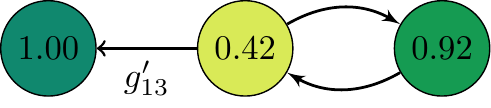}
\caption{Alternating effect of $\alpha$ is illustrated by increasing $g_{13}=0.2$ to $g_{13}'=0.4$ (bottom left to bottom right) in this network (Example \ref{ex:alt-alpha}). Numbers inside nodes at the bottom networks indicate efforts at the Nash equilibrium.}
\label{fig:3-node}
\end{figure}

\begin{example}[Alternating effect of $\alpha$]\label{ex:alt-alpha}
Consider the three node network of Fig. \ref{fig:3-node}, and a public good provision game of strategic substitutes (i.e., $g_{12},g_{13},g_{21}>0$) played on this network. Set $g_{12}=g_{21}=0.2$. Let agents' payoffs be given by: 
\begin{align*}
u_i (\mathbf{x}; \mathbf{G}) = 1-\exp(-x_i-\sum_{j\neq i} g_{ij}x_j) - \frac{1}{e}x_i~.
\end{align*}
Consider the edge between agents 1 and 3. Assume we increase the weight $g_{13}$, and want to know how this change affects the efforts of the agents at the Nash equilibrium. The results are given in the bottom two networks of Fig. \ref{fig:3-node}, and can be explained as follows. 

$\bullet$ Agent 1: the edge $1\rightarrow 3$ is on all the outgoing walks of \emph{odd} length from node 1. Increasing $g_{13}$ increases the weights of these walks. However, given $\alpha=-1$, each walk weight is multiplied by $(-1)^{2k+1}=-1$ (this is the alternating effect induced by $\alpha$). Therefore, the increase in $g_{13}$ should negatively affect agent $1$'s effort decision, leading her to decrease her effort levels in response. 

$\bullet$ Agent 2: the edge $1\rightarrow 3$ is on (some of) the outgoing walks of \emph{even} length from node 1. Increasing $g_{13}$ increases the weights of these walks. Given $\alpha=-1$, each walk weight is multiplied by $(-1)^{2k}=1$. Therefore, the increase in $g_{13}$ should positively affect agent $2$'s effort decision, leading her to increase her effort levels in response. 

$\bullet$ Agent 3: we are changing the weight of an \emph{incoming} edge to agent 3. By Theorem \ref{t:cent-eff}, only outgoing edges and walks affect the agent's effort decisions at the Nash equilibrium. Therefore, we expect agent 3's effort to remain unchanged. 
\end{example}

\begin{figure}
\centering
\includegraphics{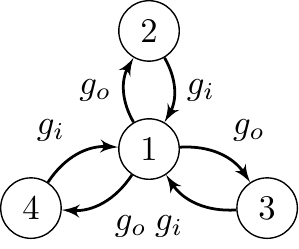}
\\
\includegraphics{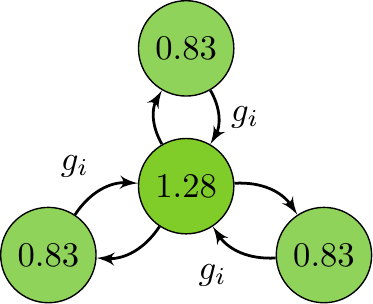}
\hspace{0.1in}
\includegraphics{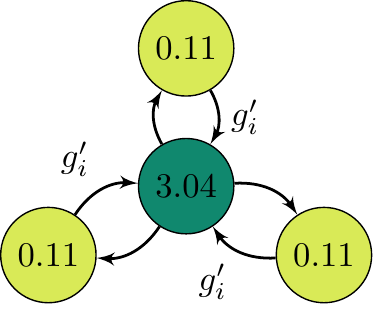}
\caption{Effect of incoming edges on perceived costs is illustrated by changing $g_{i}=0.2$ to $g_{i}'=0.3$ (bottom left to bottom right) in this network (Example \ref{ex:perceived}). Numbers inside nodes indicate efforts exerted at the socially optimal outcome.}
\label{fig:4-node}
\end{figure}

\vspace{0.1in}
\begin{example}[Effects of incoming edges and perceived costs]\label{ex:perceived}
Consider the 4 agent network of Fig. \ref{fig:4-node}. Agents' payoffs are given by:
\begin{align*}
u_1 (\mathbf{x}; \mathbf{G}) &= 1-\exp(-x_1-g_{o}\sum_{j\neq i} x_j) - \tfrac{1}{e}x_1~,\\
u_k (\mathbf{x}; \mathbf{G}) &= 1-\exp(-x_k-g_{i}x_1) - \tfrac{1}{e}x_k~, ~k\neq 1~.
\end{align*}
We consider the socially optimal effort profile in this network; i.e., $\mathbf{x}^* := \arg\max_{\mathbf{x}\geq 0} \sum_i u_i(\mathbf{x})$. This corresponds to a Pareto efficient solution of \eqref{eq:pe-lambda} with weights $\boldsymbol{\lambda}=\mathbf{1}$. Thus, according to Theorem 2, the vector of perceived costs of agents at this outcome is given by $(\mathbf{I} + \mathbf{G}^T)^{-1}\mathbf{c}$. 

Fix $g_o=0.2$. To illustrate the effect of incoming edges on agents' perceived costs, and consequently their efforts, we increase $g_i$ from $0.2$ to $0.3$. The vector of perceived costs of agents will change from $[0.17,0.33,0.33,0.33]$ to $[0.04,0.36,0.36,0.36]$. Therefore, the perceived cost of agent 1 (the center) decreases considerably when her influence on others increases, leading her to exert higher effort as a result. Furthermore, as the center invests more, the leaves now have an incentive to decrease their investment (alternating effect of $\alpha$). These effects combined lead the center (leaves) to exert higher (lower) effort when the weight of incoming edges, $g_i$, increases. 
\end{example}

\section{Extension to Coalitions}\label{sec:coalitions}
In this section, we extend the results of Section \ref{sec:cent} to effort profiles that emerge when agents belong to different coalitions. For this analysis, we distinguish between \emph{excludable} and \emph{non-excludable} public goods. With excludable goods, each coalition may choose to exclude other coalitions from experiencing the externalities of its produced good. If this is the case, each coalition can be studied in isolation, and therefore the results of the previous sections will be directly applicable. For non-excludable goods on the other hand, such separation is not possible; each coalition needs to further account for the externalities from and on other coalitions. 
Throughout this section, we are interested in the provision of such non-excludable goods. We do not explicitly model coalition formation or stability; we assume each coalition has emerged through either collaboration or appropriate incentive mechanisms. 
We present a centrality-effort connection, and the corresponding intuition, for the effort profiles emerging as the result of strategic interactions of such coalitions. 

\subsection{Semi-cooperative equilibrium}
Let agents form $K$ coalitions, denoted by the collection of disjoint sets $\mathcal{C}:=\{\mathcal{C}_1, \ldots, \mathcal{C}_K\}$, such that $\mathcal{C}_1 \cup \ldots \cup \mathcal{C}_K=\mathcal{N}$. We refer to $\mathcal{C}$ as the coalition partition. Individual agents are allowed to form their own one-member coalition. 
The effort profile emerging from the interactions of these coalitions is affected by both \emph{intragroup} and \emph{intergroup} decisions. 

Intragroup decisions refer to those adopted within each coalition. Specifically, we assume that the members within a coalition $\mathcal{C}_i$ agree (either cooperatively or through the implementation of an incentive mechanism) on a vector of welfare weights $\boldsymbol{\lambda}^i:=\{\lambda^i_k, \text{ for } k\in \mathcal{C}_i\}$, and implement the corresponding Pareto efficient solution in \eqref{eq:pe-lambda}; i.e., 
\begin{align}
\mathbf{\bar{x}}^{\boldsymbol{\lambda}^i}_{\mathcal{C}_i} = \arg\max_{\mathbf{x}_{\mathcal{C}_i}\geq \mathbf{0}} ~~ \sum_{k \in \mathcal{C}_i} \lambda_k^{i}u_k(\mathbf{x}_{\mathcal{C}_i}, \mathbf{x}_{\mathcal{N}\backslash \mathcal{C}_i})~, 
\label{eq:intragroup}
\end{align}
where $\mathbf{x}_{\mathcal{N}\backslash \mathcal{C}_i}$ denotes the efforts of agents outside the coalition. The profile $\mathbf{\bar{x}}^{\boldsymbol{\lambda}^i}_{\mathcal{C}_i}$ is therefore a Pareto efficient effort profile with weights ${\boldsymbol{\lambda}^i}$ for the agents in $\mathcal{C}_i$. 

At the intergroup level, each coalition is viewed as a \emph{super-agent}, playing a non-cooperative game with other coalitions/super-agents, and best-responding to their decisions. The resulting equilibrium effort profile $\mathbf{\bar{x}}^{\boldsymbol{\lambda}}_{\mathcal{C}}:=\left(\mathbf{\bar{x}}^{\boldsymbol{\lambda}^1}_{\mathcal{C}_1}, \ldots, \mathbf{\bar{x}}^{\boldsymbol{\lambda}^K}_{\mathcal{C}_K} \right)$ is the Nash equilibrium among these super-agents; i.e., a solution to the system of equations determined by \eqref{eq:intragroup}. We refer to $\mathbf{\bar{x}}^{\boldsymbol{\lambda}}_{\mathcal{C}}$ as a \emph{semi-cooperative} equilibrium for coalition partition $\mathcal{C}$ with weights $\boldsymbol{\lambda}$.\footnote{A semi-cooperative equilibrium is an ``equilibrium'' in the sense that, assuming binding coalition memberships, the effort profile resulting from intergroup interactions is the fixed-point of a best-response mapping. It has the limitation that it does not preclude the possibility of agents moving to other coalitions if the memberships are not binding or appropriately incentivized.}

Similar to the characterization of interior Pareto efficient outcomes in Section \ref{sec:lcp-form}, the problem of characterizing interior semi-cooperative equilibria can be formulated as an LCP. Assume agents are indexed in an order consistent with the index of their coalition memberships. Also, to simplify notation, denote the semi-cooperative equilibrium by $\mathbf{\bar{x}}$; dependence on the coalition partition ${\mathcal{C}}$ and the weights ${\boldsymbol{\lambda}}$ is implied. Then, the first order condition on \eqref{eq:intragroup} with respect to $x_i, ~ i\in\mathcal{C}_i$, implies that at the interior Pareto efficient solution, the following should hold: 
\begin{align*}
& \lambda_i b_i'(\bar{x}_i+ \sum_{j\in \mathcal{N}_i} g_{ij}\bar{x}_j) + \sum_{k\in \mathcal{C}_i, \text{ s.t. } i\in \mathcal{N}_k} \lambda_k g_{ki} b_k'(\bar{x}_k + \sum_{j\in \mathcal{N}_k} g_{kj}\bar{x}_j) = \lambda_i c_i,~ \forall i~.
\end{align*}

Define $q_i^{\mathcal{C}, \boldsymbol{\lambda}}$ as the effort levels at which:
\[b_i'(q_i^{\mathcal{C}, \boldsymbol{\lambda}})=\left((\mathbf{I} + \boldsymbol{\Lambda}^{-1}\mathbf{G}_{\mathcal{C}}^T\boldsymbol{\Lambda})^{-1} \mathbf{c} \right)_i~.\]
Note that the only difference of these efforts with the $q_i^{\boldsymbol{\lambda}}$ defined for Pareto efficient outcomes is in the \emph{coalition-modified} dependence matrix $\mathbf{G}_{\mathcal{C}}$, which is defined as follows: for each row $k$ corresponding to an agent in coalition $\mathcal{C}_i$, set the entries $g_{kl}, \forall l\notin \mathcal{C}_i$ to zero. This matrix is therefore equivalent to the dependence matrix of a network obtained by removing all edges between coalitions.

Then, finding an interior semi-cooperative equilibrium $\mathbf{\bar{x}}$ is equivalent to finding a solution with $\mathbf{w}=\mathbf{0}$ to the following LCP: 
\begin{align}
\mathbf{w} - (\mathbf{I} + \mathbf{G})\mathbf{x} = -\mathbf{{q}^{\mathcal{C},\boldsymbol{\lambda}}}~,\notag\\
\mathbf{w}\succeq \mathbf{0}~, ~~ \mathbf{x}\succeq \mathbf{0}~,\notag\\
\mathbf{w}^T\mathbf{x} = 0~.
\label{eq:lcp-coop}
\end{align}

Using a similar procedure as Section \ref{sec:interior-exist}, such profile exists under the following condition. 

\vspace{0.16in}
\begin{theorem}[Existence of interior semi-cooperative equilibria]\label{t:interior-coop}
The public good provision game has an interior semi-cooperative equilibrium $\mathbf{\bar{x}}^{\boldsymbol{\lambda}}_{\mathcal{C}}$ if and only if the corresponding $\mathbf{{q}^{\mathcal{C},\boldsymbol{\lambda}}}$ is in the positive cone generated by the columns of $\mathbf{I}+\mathbf{G}$. 
\end{theorem}

\subsection{A centrality-effort connection} \label{sec:cent-coalitions}

We now present a centrality-effort characterization of interior semi-cooperative equilibria. 

\vspace{0.1in}
\begin{theorem}[Centrality-effort connection for semi-cooperative equilibria]\label{t:cent-eff-coop} 
Consider an interior semi-cooperative equilibrium $\mathbf{\bar{x}}^{\boldsymbol{\lambda}}_{\mathcal{C}}$. Then, 
\[\mathbf{\bar{x}}^{\boldsymbol{\lambda}}_{\mathcal{C}} = c_\text{alpha} \left(\mathbf{G}, -1, \mathbf{q}^{\mathcal{C},\boldsymbol{\lambda}}\right)~,\]
where $\mathbf{q}^{\mathcal{C}, \boldsymbol{\lambda}}$ is such that $b'_i(q^{\mathcal{C},\boldsymbol{\lambda}}_i)=c_{\text{alpha}, i}(\boldsymbol{\Lambda}^{-1}\mathbf{G}_{\mathcal{C}}^T\boldsymbol{\Lambda}, -1, \mathbf{c})$. 
\end{theorem}
\vspace{0.1in}
\begin{IEEEproof}
An interior semi-cooperative equilibrium for coalition partition $\mathcal{C}$ and weights $\boldsymbol{\lambda}$ is a solution to LCP \eqref{eq:lcp-coop} with $\mathbf{w}=\mathbf{0}$; i.e.,
\begin{align*}
(\mathbf{I}+\mathbf{G})\mathbf{x} = \mathbf{q}^{\mathcal{C},\boldsymbol{\lambda}}, ~~ \mathbf{x}\succeq \mathbf{0}~.
\end{align*}
Therefore, when such solution exists, $\mathbf{\bar{x}}^{\boldsymbol{\lambda}}_{\mathcal{C}} = (\mathbf{I}+\mathbf{G})^{-1}\mathbf{q}^{\mathcal{C},\boldsymbol{\lambda}}$. Also, by definition, we know that $\mathbf{q}^{\mathcal{C},\boldsymbol{\lambda}}$ satisfies $b_i'(q_i^{\mathcal{C}, \boldsymbol{\lambda}})=\left((\mathbf{I} + \boldsymbol{\Lambda}^{-1}\mathbf{G}_{\mathcal{C}}^T\boldsymbol{\Lambda})^{-1} \mathbf{c} \right)_i.$ Comparing these expressions with \eqref{eq:alpha-cent} establishes the connection. 
\end{IEEEproof}
\vspace{0.1in}

The implications of the centrality-effort connection on effects of incoming and outgoing edges and the alternating effect induced by $\alpha=-1$ are applicable to the characterization of Theorem \ref{t:cent-eff-coop} as well. The main difference resulting from the formation of coalitions can be explained as follows. 

\vspace{0.1in}
\textbf{The effect of coalitions -- benefiting from dependencies and ignoring influences:}   with non-excludable goods, an agent can benefit from the externalities of the effort exerted by her neighbor, whether or not that neighbor is a member of the agent's coalition. Consequently, the alpha-centralities (i.e, efforts) of agents are calculated on the full network of dependencies $\mathbf{G}$. 

However, recall that the perceived costs of each agent are affected by {the influence of the agent on those with who} she is cooperating to implement a Pareto efficient effort profile. 
The agents in a coalition account for their influences on others in their group, but disregard their influence on all other agents. Therefore, agents' perceived costs $c_{\text{alpha}, i}(\boldsymbol{\Lambda}^{-1}\mathbf{G}_{\mathcal{C}}^T\boldsymbol{\Lambda}, -1, \mathbf{c})$, while again evaluated on the network of influences (i.e., the transpose of the dependence matrix), are now evaluated on a coalition-modified matrix of influences $\mathbf{G}_{\mathcal{C}}^T$. In other words, when determining their perceived costs, agents act as if the dependence network is one in which all edges between coalitions are removed.

\section{Conclusion}\label{sec:conclusion}

We studied the provision of public goods on a network of strategic agents. We identified a necessary and sufficient condition on the dependence matrix of the network that guarantees the uniqueness of the Nash equilibrium in these games. Our condition unifies (and strengthens) existing results in the literature. We also identified necessary and sufficient conditions for existence of Nash equilibria in subclasses of games that lie at the two extremes of our model; namely games of strategic complements and games of strategic substitutes. An interesting direction of future work is to identify similar conditions for a general model of games on networks, and in particular, games with non-linear best replies. 

We further presented a graph theoretical characterization of different interior effort outcomes, namely Nash equilibria, Pareto efficient outcomes, and semi-cooperative equilibria, in terms of agents' alpha-centralities in their dependence network. Using this characterization, we were able to identify the effects of incoming edges, outgoing edges, and coalitions, as well as an alternating effect over walks of different length in the network. As part of our future work, we are interested in using this connection for conducting comparative statics (e.g., the effects of adding/removing links), as well as for the design of targeted tax/subsidy policies that can incentivize the improved provision of the public good. 

\section*{Acknowledgment}
The authors would like to thank Hamidreza Tavafoghi for useful discussions and comments on earlier drafts of this work. This work is supported by the Department of Homeland Security (DHS) Science and Technology Directorate, Homeland Security Advanced Research Projects Agency (HSARPA), Cyber Security Division (DHS S\&T/HSARPA/CSD), BAA 11-02 via contract number HSHQDC-13-C-B0015. 

\bibliographystyle{IEEEtran}
\bibliography{IEEEabrv,networkgames}

\end{document}